\documentclass[twocolumn,showpacs,preprintnumbers,amsmath,amssymb]{revtex4}

\usepackage{graphicx}
\usepackage{dcolumn}
\usepackage{bm}
\newcommand{\etal}{\emph{et al.}}

\begin{document}
\title{Phase coherence and the Nernst effect at magic angles in organic conductors }
\author{N. P. Ong, Weida Wu, P. M. Chaikin, and P. W. Anderson}
\affiliation{Department of Physics, Princeton University, Princeton, N.J. 08544, U. S. A.
}

\date{\today}

\begin{abstract}
A giant Nernst signal was recently observed for fields near
crystallographic directions in (TMTSF)$_2$PF$_6$. Such large
Nernst signals are most naturally associated with the motion of
pancake vortices.  We propose a model in which phase coherence is
destroyed throughout the sample except in planes closely aligned
with the applied field $\bf H$. A small tilt above or below the plane
changes the direction and density of the penetrating vortices and
leads to a Nernst signal that varies with the tilt angle of $\bf
H$ as observed. The resistance notches at magic angles are understood 
in terms of flux-flow dissipation from field-induced vortices.
\end{abstract}
\maketitle                   

\begin{figure}              
\includegraphics[width=8cm]{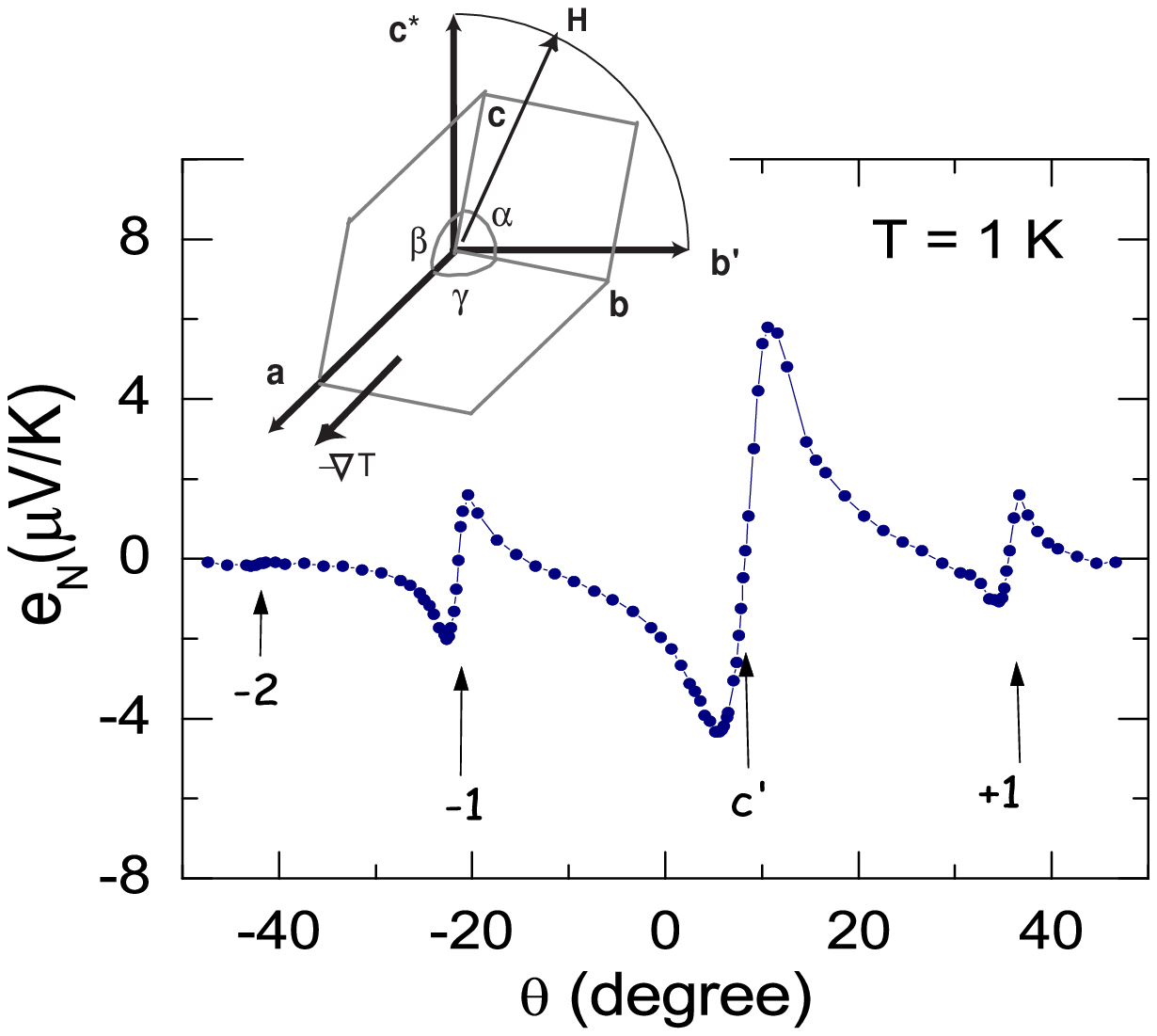}
\caption{\label{Ntheta}  The Nernst signal
$e_N$ vs. tilt angle $\theta$ in (TMTSF)$_2$PF$_6$ measured
with $H$ = 7.5 T, $T$ = 1 K and 
$P$ = 10 kbar (here $\vec{-\nabla T}||\bf c$ and ${\bf E}_N||{\bf a}$).  $e_N$ changes sign at the magic angles
$\theta_{-L1}$, $\theta_c$ and $\theta_{L1}$.  The inset shows the standard experimental 
arrangement in which $\vec{-\nabla T}||\bf \hat{x}||a$, $\bf \hat{y}||b$ and $\bf \hat{z}||c^*$, 
with reciprocal vectors $\bf b^*|| c\times a$ and $\bf c^*|| a\times b$.  
The field tilt-angle $\theta$ is measured from the direction $\bf c^*$.  
}
\end{figure}

At low temperatures, the quasi-one dimensional organic conductor
(TMTSF)$_2$PF$_6$  displays a rich assortment
of  electronic phases, from spin-density-wave (SDW) to
superconductivity to the field-induced spin-density-wave
- quantum Hall effect (FISDW), as the applied  pressure and magnetic
field are varied~\cite{Kang,Chaikin,Danner,Chash}.  For example,
at a fixed pressure $P$ = 9 kbar, (TMTSF)$_2$PF$_6$ is a 
superconductor in zero field at temperatures $T$ below $T_c$ = 1.2
K.  In a moderate field $\bf H$, the zero-resistance state
is destroyed in favor of a putative `metallic' state with
large angular magnetoresistance.   At larger fields, a cascade of phase
transitions occurs to semi-metallic and at the highest fields an insulating FISDW state.  A 
striking feature in the metallic state is the existence of sharp
notch anomalies in the resistance when $\bf H$ is aligned with the
`magic  angles'~\cite{Kang,Chaikin,Danner,Chash}.  Lebed \cite{Lebed} originally predicted magic-angle 
effects with resistance peaks.  However, resistance minima are observed instead. 
The origin of the magic-angle anomalies is an open problem despite a large
number of proposed models~\cite{models,Strong}.  Recently, Wu, Lee
and Chaikin (WLC)~\cite{Wu} uncovered a remarkable angular Nernst
effect in the metallic state of (TMTSF)$_2$PF$_6$.  The
unusual angular dependence and the large magnitude of the Nernst
signal (Fig. \ref{Ntheta}) seem incompatible with the conventional transport 
theories.  Here we show that a large resonant Nernst signal at the
magic-angles can result from phase slip and the partial restoration of
long-range  phase coherence of the superconducting pairing
condensate when $\bf H$ is exactly aligned with a set of crystal
planes.

Conventionally, the Nernst effect corresponds to the appearance
of a transverse electric field $\bf E_N$ that is antisymmetric in both
$\bf H$ and the applied temperature gradient $-\vec{\nabla T}$, 
i.e. ${\bf E}_N\sim {\bf H}\times\vec{\nabla T}$
(for e.g. $\bf E_N||\hat{z}$ if $-\vec{\nabla T}||\mathbf{\hat{x}}$ and
$\bf H||\hat{y}$)~\cite{Kim,Wang1}.   We refer to 
$e_N \equiv E_N/|\vec{\nabla T}|$ as the Nernst signal. 
In the  experiment of WLC~\cite{Wu}, $\bf H$ is rotated in 
the $b^*c^*$ plane (the plane 
normal to $\bf a$; see Fig. \ref{Ntheta} inset).  In the metallic
state of (TMTSF)$_2$PF$_6$, they observed that, as $\theta$ is
varied, the curve of $e_N$ vs. $\theta$ is comprised of a series of
sharp resonant-like curves centered at the magic angles (Fig. \ref{Ntheta}).  

Several features of the Nernst experiment are noteworthy.  As noted, in conventional Nernst 
experiments, $E_N$ changes its sign like the cross-product 
${\bf H}\times\vec{\nabla T}$.  By contrast, the sign-change in $e_N$ at 
the magic angles in Fig. \ref{Ntheta} 
occurs even though neither ${\bf H}$ nor $\vec{\nabla} T$ changes 
sign as $\theta$ crosses a magic angle.  
Here, the sign of $e_N$ reflects rather the component of $\bf H$ normal 
to a crystal plane, e.g. $\bf H\times c'$ (a similar dependence 
of the magnetoresistance on $\bf H\times c'$ was previously noted).  
Equally puzzling, the peak magnitude of $e_N$ at $T \le$ 1 K is $10^3$-$10^5$ 
times larger than that derived from quasiparticle currents calculated from 
the band structure of (TMTSF)$_2$PF$_6$.  Moreover, the peak signal 
increases rapidly as $T$ decreases from 4 to 0.5 K, whereas a quasiparticle signal 
should decrease monotonically.  Lastly, the large Nernst signal occurs
in the face of an undetectably small thermopower $S$.  This is anomalous for charge carriers 
because their drift velocity component $||(-\vec{\nabla T})$ is generally much larger than the
transverse component produced by the Lorentz force (i.e. $S\gg e_N$)~\cite{Wang1}.  
By contrast, vortex flow $||(-\vec{\nabla T})$ produces an $E$ field 
that is predominantly transverse.  Hence $e_N\gg S$ (as observed) is a strong clue 
that the Nernst signal originates from vortex flow rather than charge carriers.

\begin{figure}[h]           
\includegraphics[width=6cm]{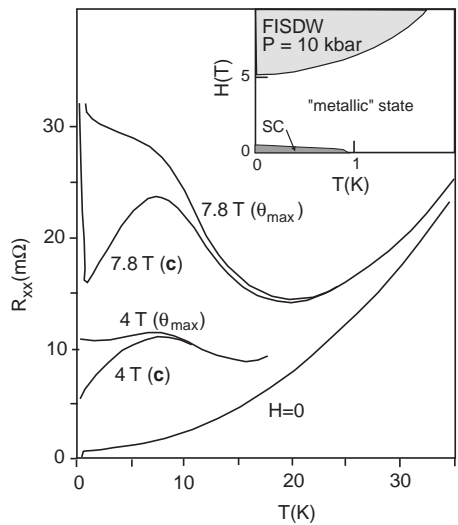}
\caption{\label{RvsT}  The $T$ dependence of $R_{xx}$ measured at $H$ = 0, 4, and 7.8 T.  Curves with $\bf H || c'$ are compared with those measured 
with $\bf H$ in a local-maximum direction ($\theta_{max}$) [adapted from Ref. \cite{Chash}].  The inset is a schematic of the phase diagram in the 
$H$-$T$ plane at $P$ = 10 kbar.  Superconductivity (SC) with long-range phase coherence is confined to the darker shaded region.  We propose that 
$|\hat{\psi}({\bf r})|$ survives in the chains deep into the metallic state. 
}
\end{figure}

For the pressure $P$ (7-10 kbar) and field (4-8 T) employed, (TMTSF)$_2$PF$_6$
is in the `metallic' state, which falls between the superconducting state (in which long-range phase
coherence is fully established) and the FISDW state (Fig. \ref{RvsT}, inset).   We propose that, over 
large regions of the `metallic' state, the Cooper-pairing amplitude $|\hat{\psi}({\bf r})|$ remains large within 
each chain along $\bf a$ but long-range phase coherence is absent for general field directions.  
The Nernst signal arises from phase slippage caused by vortex flow in these 
planes as $\bf H$ tilts off alignment.

In a type II superconductor with extreme anisotropy and small
superfluid density $\rho_s$ (due to low carrier density),
superfluidity (long-range phase  coherence) is easily destroyed
by strong non-Gaussian fluctuations of the phase $\varphi(\bf r)$
in zero field even if $|\hat{\psi}|$ remains 
large~\cite{Emery}.   In finite field, loss of long-range phase
coherence at the vortex solid-to-liquid transition leads to a
strongly dissipative vortex-liquid  state.  The sharp increase
in resistance at the melting line is often mistaken for the upper
critical field.  A series of 
experiments~\cite{Xu,Wang1,Wang2,Wang3,Capan,Ong} show that the vortex
liquid state in cuprates is indistinguishable from the
normal state by  resistance measurements, but may be detected
in a Nernst experiment.  The persistence of the vortex liquid
state to regions of the phase diagram high  above the Meissner
state boundary may be easily missed using resistivity data alone.

\begin{figure}              
\includegraphics[width=6cm]{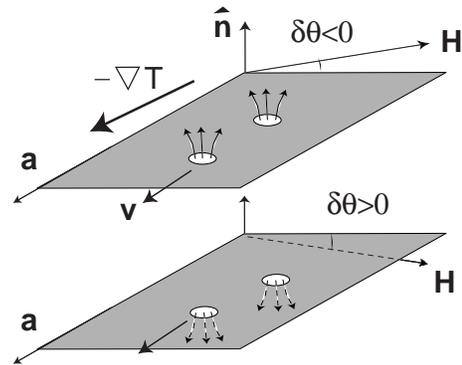}
\caption{\label{vortex} Sketch of pancake vortices (circles with directed flux lines) created by $\bf H$ when it is near alignment with a plane (shaded 
surface).  The density of vortices equals $\bf B\cdot \hat{n}$.  Phase slippage from the flow of vortices parallel to $-\vec{\nabla T}$ induces a large Nernst 
signal $e_N$.  As $\bf H$ tilts below the plane, $\bf B\cdot \hat{n}$ and the vorticity change sign (lower sketch).  Hence $e_N$ 
also changes sign.
}
\end{figure}

\begin{figure}              
\includegraphics[width=6cm]{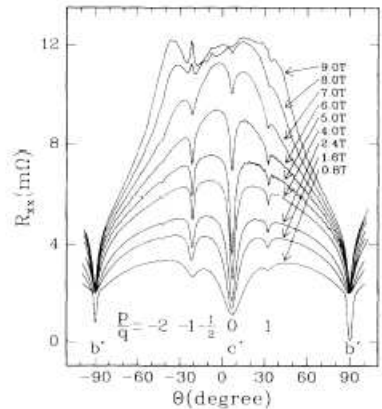}
\caption{\label{Rtheta}  The angular dependence of $R_{xx}$ in (TMTSF)$_2$PF$_6$ at 0.5 K with $P$ = 10 kbar, at selected fields.  Sharp notches in 
$R_{xx}$ occur at the magic angles (Ref. \cite{Kang}).
}
\end{figure}
Our starting assumption is that, in
the `metallic' state of (TMTSF)$_2$PF$_6$, loss of phase
coherence arises from highly mobile 2D vortices living
on the principal crystal planes  $ab$, $ac$ or $a(c\pm b)$.  
[The transfer integrals $t_i$ in the 3 principal bond directions are in the
ratio $t_a:t_b:t_c = 1:0.1:0.003$~\cite{Kang}.  Accordingly, the
pairing strength is largest along  $\bf a$, but progressively
weaker along the bond directions $\bf b$, $\bf c$, $\bf (c-b)$ and
$\bf (c+b)$~\cite{Chaikin}.] In a field $\bf H$ at finite $T$, the vortex population is
comprised of thermally generated  and field-induced vortices
(Fig. \ref{vortex}).  Both the vorticity and population of the
field-induced vortices are determined by $B_n \equiv\bf
\hat{n}\cdot  B$ where $\bf \hat{n}$ is the unit vector normal
to a principal plane and $\bf B$ the induction field.  If $\bf H$
is exactly aligned with the plane, we have only  thermally
generated pairs of `up' and `down' vortices.  (Planes not aligned with 
H are saturated with vortices to the extent that they have lost phase coherence.) 
Tilting $\bf H$ slightly out of the plane (Fig. \ref{vortex}) leads to a steep
increase in the field-induced  population equal to $\bf B\cdot \hat{n}$, 
and an increase in the resistance $R_{xx}$ as displayed in Fig. \ref{RvsT}.  

The flow of vortices parallel to $-\nabla T$ leads naturally to a Nernst signal that changes sign with
$\bf B\cdot \hat{n}$ as observed by WLC~\cite{Wu}.
We consider the case when $\bf H$ is close to alignment with the
$ac$ plane (i.e. $\bf H||c'$ at the magic angle $\theta_c$, Fig. \ref{vortex}).  The gradient $-\vec{\nabla T}||{\bf a}$
produces a vortex flow in this  plane with velocity $\bf v||a$.
The passage of each vortex across a line $\perp\bf a$
causes a phase slip of $2\pi$.  By the Josephson equation, the
rate of phase slippage $\dot{\varphi}$ leads to an
electro-chemical potential difference between the two ends of the line $V_J =
\hbar\dot{\varphi}/2e$, which  translates to a Nernst electric field 
${\bf E}_N = B_n\bf \hat{n}\times v$.   Clearly, as we tilt
$\bf H$ from just above the plane to just below, $E_N$ changes 
sign with $B_n$ (Fig. \ref{vortex}).

We estimate the magnitude of $e_N$ as follows.  The gradient exerts
a line force $s_{\phi}(-\vec{\nabla T})$ on each vortex, where $s_{\phi}$
is the `transport line entropy'~\cite{Kim,Wang1}.  Equating this to the frictional damping
force $\eta{\bf v}$, we have $v = s_{\phi}(-\vec{\nabla T})/\eta$, with $\eta$
the damping parameter.   With $B_n\sim (\theta-\theta_c) B$, we have for the Nernst electric field along $\bf c$
\begin{equation}
E_N =  (\theta-\theta_c) B\frac{s_{\phi}}{\eta}\left[-\vec{\nabla T} \right], \quad\quad (\theta\sim \theta_c).
\label{EN}
\end{equation}
In effect, as $\delta\theta = \theta-\theta_c$ changes sign, the
field-induced vortices decrease in number to zero, and then
reappear with the opposite circulation.  Since ${\bf v}||(-\nabla
T)$ is unchanged, the Nernst signal 
changes in sign.

At small $|\delta\theta|$, the resistance and $E_N$ both vary linearly
with $B_n$.  When these two quantities scale in the same way with
field, their magnitudes  are related through the damping
parameter $\eta$.   By Eqs. \ref{EN} and \ref{rhoc} (see below),
we have $\partial e_N/\partial B_n = s_\phi/\eta$ and 
$\partial \rho_c/\partial B_n = \phi_0/\eta$, respectively with $\phi_0$ the flux quantum.  We
adopt the value $s_\phi \sim 1\  k_B$ per 2D vortex inferred
from experiments on  cuprates~\cite{Wang2}.   From the results
of Wu \etal at $H$ = 7 T and $T$ = 1 K~\cite{Wu}, $\rho_c$ is $\simeq \ 3\   \Omega$cm at
$\theta_c$, which implies $\partial \rho_c/\partial \theta =
3.8\times 10^{-6} \ \Omega$m/deg.  Using this and $s_\phi/\phi_0
\simeq$ 9  A/Km, we find $\partial e_N/\partial \theta \simeq
30 \ \mu \rm{V( K deg)^{-1}}$.  We estimate a peak value
$e_N \simeq 200\ \mu$V/K when ${\bf E}_N$ is measured along $\bf c$ in  
a gradient $||{\bf a}$ consistent with measurements.  If, however, ${\bf E}_N$ is measured along $\bf a$ with 
$\vec{\nabla} T ||\bf c$ (as in the main panel in Fig. \ref{Ntheta}), we should scale with the resistive notch 
in $\rho_a \sim 20 \mu \Omega$cm.  This gives a peak 
$e_N \simeq 2\ \mu$V/K, which is again consistent 
with experiment.  As noted, calculations~\cite{Wu} of the
Nernst signal produced by charge carriers are too small by 
a factor of $10^3$-$10^5$ at 1 K.

When $|\delta\theta|$ exceeds $15^{\rm o}$, $\rho_a$ saturates to the
 normal-state dome-shaped background as phase coherence is reduced to
 short length scales.  The Nernst signal reaches a peak near this tilt
angle, and falls monotonically until $\theta$ reaches the next magic angle.
(The  peaking of the Nernst signal at a field close to where the resistance saturates 
is closely similar to what is observed in a recent experiment on  cuprates~\cite{Wang3}.)  
The concatenation of such dispersion-like curves at successive magic angles 
accounts for the curve in Fig. \ref{Ntheta}.

As the vortex velocity $\bf v$ is nearly parallel with $-\vec{\nabla T}$, the component of ${\bf E}_N$ along $\bf a$ 
(which is detected as a thermopower signal) is negligible.  This readily explains why $S$ remains unresolvably small even 
as $e_N$ increases to very large values with decreasing $T$. 

The $T$ dependence of the Nernst signal is also consistent with a
vortex origin.  As $T$ decreases from 4 K to 0.5 K, $e_N$
increases dramatically by  $\sim$10 at $\theta_c$ (and even
more steeply at $\theta_{\pm L1}$)~\cite{Wu}.  This behavior is
incompatible with charge carriers  (for which both $e_N$ and $S$ 
decrease monotonically with $T$).  For a vortex liquid, the $T$ dependence of $e_N(T,H)$ is 
most well-studied in underdoped $\rm La_{2-x}Sr_xCuO_4$.  There, as $T$ decreases from 80 K in fixed $H$, $e_N$ rises 
rapidly, continuing to do so below the zero-field critical temperature $T_{c0}$ = 28 K (Fig. 1 of Ref. \cite{Ong}).  
At lower $T$, $e_N(T,H)$ attains a prominent maximum at $T_p$, and then
decreases rapidly to zero as the vortex-to-solid melting temperature $T_M$ is approached ($T_M<T_p<T_{c0}$).  
We expect that, near each magic-angle direction in (TMTSF)$_2$PF$_6$, $e_N$ 
eventually rises to a peak before decreasing rapidly to zero as the 2D vortices approach solidification below 0.5 K.  
As $t_c > t_{c\pm b}$, the peak should occur at a higher $T$ with $\bf H||c$ (at $\theta_{c}$) than with $\bf H||(c\pm b)$ 
($\theta_{\pm L1}$).

The 2D vortex model may also explain why sharp resistance minima are observed at the 
magic angles (Fig. \ref{Rtheta}).  We assume $\bf H$ is precisely aligned with 
the $ac$ plane.  Within the plane, phase coherence of the superconducting order parameter $|\psi({\bf r})|\mathrm{e}^{i\varphi({\bf r})}$ extends to the 
Kosterlitz-Thouless (KT) correlation length $\xi_+$, which measures the spacing between thermally 
generated vortices (of density $n_f$).   The area of the average phase-correlated region $\xi_+^2$ dictates the magnitude 
of the superfluid conductivity enhancement $\sigma_s$.   For an isotropic plane, Halperin and Nelson~\cite{Halperin} 
derived $\sigma_s = \sigma^{(n)}/n_f2\pi\xi^2$, with $\sigma^{(n)}$ the normal-state resistance, and $\xi$ 
the Ginzburg-Landau coherence length evaluated at $T_{c}$.  
As shown in Fig. \ref{vortex}, tilting $\bf H$ slightly away from $\theta_c$ produces 2D vortices of density 
$n_B = |H\sin\delta\theta| /\phi_0$ in that  plane.  
The mobile field-induced vortices dramatically shrink the phase-correlated area, and result in 
a steep suppression of $\sigma_{c}^{(s)}$.  Hence, for small $\delta\theta$, we may write for the $a$-axis resistivity
\begin{equation}
\rho^{(s)}_a = \rho^{(n)}_a 2\pi (\xi_a\xi_c) \left( n_f + \frac{H}{\phi_0}|\theta-\theta_c| \right).
\label{rhoc}
\end{equation}

The singular behavior $\delta\rho^{(s)}_a\sim |\delta\theta|$ in Eq. \ref{rhoc} implies 
that the resistivity displays a sharp notch when $\theta$ is close to  $\theta_c$ as observed.  Let us consider the curve of $R_{xx}$ (or $\rho_a$) vs. 
$\theta$ (Fig. \ref{Rtheta}).  As $\bf H$ is rotated in the plane $b^*c^*$
normal to $\bf a$, $R_{xx}$  displays a series of sharp
notches~\cite{Kang}.  At $\theta_c \simeq 4^\mathrm{o}$, $n_B$ vanishes in the
$ac$ plane.  The restoration of long-range phase
correlation in $ac$ leads to a large rise in conductance observed
as the steep decrease in $R_{xx}$ in both Figs. \ref{RvsT} and \ref{Rtheta}.  
As $\bf H$ deviates from
$\theta_c$, $\rho^{(s)}_c$ increases steeply with $n_B\sim H
|\delta\theta|$.  When $n_B$ gets so large that  phase
coherence in the $ac$ plane is reduced to very short length
scales, $R_{xx}$ reverts to the dome-shaped background
($|\delta\theta| > 15^{\rm o}$) [Fig. \ref{Rtheta}].   Similarly, the weak notch at
the magic angle $\theta_{L1}\simeq\ 32^\mathrm{o}$ occurs when
$\bf H||(c+b)$.  The conductance increase is relatively  modest
because of the weak Josephson coupling in the plane $a(c+b)$.
Finally, when $\bf H||b$ (at $\theta_b$), long-range phase
coherence is restored  in the strongly coupled plane $ab$,
resulting in a deep notch in $R_{xx}$.  (The notch at
$\theta_{L1}$ is slightly weaker than that at $\theta_{-L1}\simeq\
 -20^\mathrm{o}$.  The slight asymmetry reflects the slight
bond-length difference ($\bf |c+b| > |c-b|$) which implies a
larger $t$ along $\bf c-b$.)  

Implicit in our model is the surprising ability of the magnetic
field to destroy phase coherence in the planes to which it is normal
while enhancing conductivity in the planes to which it is parallel.  It is natural 
that the vortices penetrating perpendicular planes destroy superconductivity. 
It is also natural that a field applied between two highly conducting/superconducting 
planes tends to decouple them by free flowing Josephson vortices ala the cuprate 
superconductors.  However, in the present scenario, it is the {\em least} conducting 
planes which remain internally coherent and they remain decoupled from each other 
even though the interplane bandwidth (along $\bf b$) is more than an order of magnitude 
larger than the intraplane bandwidth (along $\bf c$). From the usual behavior of anisotropic 
superconductors, this magic-angle behavior seems rather counter-intuitive, 
although not precluded in principle.  

A physical picture of superconductivity in this material is that it results from interactions 
within the chains, in which the $p$-wave pairing susceptibility is enhanced by the 
same spin fluctuations which cause a Mott insulating SDW at lower pressures.
However, because of Coulomb blockade and localization effects, a strictly 1D chain is not superconducting 
as $T\rightarrow 0$.  Hence conductivity must arise from interchain coupling.  In weak fields, we 
have 3D superconductivity, but when interplane coupling is suppressed by $\bf H|| b$ or $\bf c$, 
the 2D planes become resistive (they lie above their KT transition).  The phase stiffness in the 
direction $\bf b$ is proportional to $t_b^2/t_a$ and hence also small.  Thus, the magic angles are, in a sense, 
where restoration of interchain coupling allows the current to bypass obstructions along the chains.  
In principle this mechanism is similar to that proposed in Strong \etal~\cite{Strong}.

In summary, we have shown that the angular Nernst signal in the
`metallic' phase of (TMTSF)$_2$PF$_6$ is well accounted for in a
model in which the  Cooper pairing amplitude $|\hat{\psi}|$
remains finite, but superfluidity is absent because of dominant
phase fluctuations produced by mobile vortices.   When $\bf H$
tilts away from a principal plane direction, the sharp increase in
2D vortex density leads to a notch in the resistance as well as an
angular  Nernst signal of the observed magnitude.

We acknowledge discussions with Yayu Wang, Injae Lee and Michael Naughton.  This research is supported by the U. S.
National Science  Foundation (MRSEC Grant DMR 0213706) and DMR-0243001.

\end{document}